\documentclass{article}
\usepackage[utf8]{inputenc}
\usepackage{bbm}
\usepackage{breqn}
\usepackage{dsfont}
\usepackage{comment}
\usepackage{graphicx}
\usepackage{amsmath}
\usepackage{tikz}
\usepackage[title]{appendix}

\newcommand{\prob}{\text{Prob}}
\usepackage{biblatex} 
\numberwithin{equation}{section}

\newcounter{remark}

\newcommand{\kb}[2]{\ket #1 \! \bra #2}
\newcommand{\kbp}[2]{| #1) \! ( #2|}

\newcommand{\bra}[1]{\left\langle #1 \right|}

\newcommand{\braket}[2]{\langle #1 | #2 \rangle}
\newcommand{\ket}[1]{\left| #1 \right\rangle}
\title{Lindblad evolutions, Born rule, Heralding and cloning}
\author{A. Auerbach, J. Avron, D. Gershoni, N. Tur}
\date\today

\begin{document}

\maketitle
\begin{abstract}
We discuss an apparent difficulty in computing the radiation emitted by a system undergoing Lindblad evolution. The difficulty is resolved by viewing  the problem as Born rule for conserved currents defined by the appropriate terms in the adjoint Lindbladian. 

In Heralding Alice prepares Bob's system in a state that mirrors her test.  We show that heralding is consistent with no-cloning. This follows from the observation that heralding is not a completely positive map.
\end{abstract}
\hfill
\begin{minipage}[r]{.5\textwidth}
\it{To Michael Berry with gratitude for his friendship, insights, and all that he taught us.}
\end{minipage}

\section{The problem and its resolution}
It is a common practice to model simple radiating systems, such as  atoms or a quantum dots, by a finite dimensional system, whose quantum state $\rho$ evolves according to a suitable Lindblad operator \cite{breuer,nielsen,wiseman}. The radiation field (and possibly other degrees of freedom) are modeled by suitable ``jump operators" in the Lindbladian.  
In the special case of a 2-level system  the photo-emission is found to be proportional to 
\begin{equation}\label{e:algorithm}
    Tr\Big(\rho \kb{1}{1}\Big)
\end{equation}
where $\kb{1}{1}$  projects on the excited, i.e. radiating, energy level. 

There are several difficulties with this formula. First, $\rho$ is a Markovian approximation to the  partial trace of the joint system and radiation:
\begin{equation}
    \rho=Tr_{r}\, \rho_{sr}, \quad \rho\in\mathbf{S}, \ \rho_{sr}\in\mathbf{S}\otimes \mathbf{R}
\end{equation}
As the partial trace erases the information regarding the radiation degrees of freedom, it is surprising that one can extract information on the radiation from knowledge of the reduced density matrix  of the system. 

Second, Eq.~\ref{e:algorithm} has the form of Born rule for the detecting the excited state. This is puzzling: By the rules of quantum mechanics, a measurement  prepares the system\footnote{A single shot does not reveal the quantum state prior to the measurement, which requires tomography, and hence many copies.}. Since the detection of the radiation prepared the system in the ground state the Born rule should have been  $Tr\,\big(\rho\kb{0}{0}\big)$ rather than Eq.~\ref{e:algorithm}. This, however, leads to the patently absurd result that the system in the ground state radiates. 

These difficulties suggest that the observable $\kb{1}{1}$ in Eq.~\ref{e:algorithm} has an alternate interpretation that agrees with the Born rule. This interpretation should also explains how come measurements performed on the radiation can be determined from the reduced density matrix of the system.

The protection of information comes from conservation laws. Indeed, the reduced density matrix $\rho$ carries no information on the radiation, however, there is such information  in $\dot \rho$.  The conserved quantity is the ``total excitation number" which is exact constant of motion in the rotation wave approximation. It says that the sum of the excitation and photon number is conserved. It is, of course, related to energy conservation\footnote{It is equivalent to energy conservation if one neglects the energy associated with the coupling of the systems. This is a good approximation in the long time limit, when the systems separate.}. In general, one may have to consider also other conservation laws, such as conservation of angular momentum. We discuss this in section \ref{s:am}.

Due to the conservation law of total excitation,  the (outgoing) radiation current and the (incoming) excitation current, balance the population rate of the corresponding energy level. In section \ref{s:2l}, we construct the {\em system observable}  corresponding to the contributions of radiation to the population rate. In the case of a two-level system it is given by either:  
\begin{equation}\label{e:current}
    \mu D^*(A,\kb{0}{0})= -\mu D^*(A,\kb{1}{1})
    \end{equation}
$\mu D^*$ is the radiative term in the adjoint Lindbladian $L^*$, see Eq.~\ref{e:Ls} below. $A$ is the radiation jump operator, see Eq.~\ref{e:jump}, and $\mu$ the decay rate. The two sides of the equation represent the population rate of the ground state and minus the population rate of the excited state. The formula is independent of the choice of basis for representing the jump operator $A$ and the projection $\kb{0}{0}$.

The situation in a two level system is particularly simple. Inserting the jump operator, Eq.~\ref{e:jump}, in Eq.~\ref{e:current} gives 
\begin{equation}
    \mu D^*(A,\kb{0}{0})= \mu\kb{1}{1} 
\end{equation}
The radiation current happens to be proportional to the projection on the radiating level. Eq.~\ref{e:algorithm} is therefore {\em not} the Born  rule for the detection of the excited state, but the Born rule for the contribution of radiation to the population rate of the ground state (or its minus for the excited state).

It may well be that neither   Eq.~\ref{e:current} nor its interpretation are new. But, we did not find either in standard textbooks \cite{wiseman,breuer,carmichael}. This insight seems worth repeating even if it is not new.

\section{Relaxation currents in Lindblad evolution}\label{s:2l}
For the sake of simplicity we start by considering a Lindbladian describing radiative excitation and decay of a two level system.  $\rho$ is a positive $2\times 2$ matrix.

The Lindbladian evolution is given by 
\begin{equation}
    \dot\rho=L(\rho)
    \end{equation}
where
\begin{equation}\label{e:lind}
    L(\rho)= -i [H,\rho] +\mu D(A,\rho) + \lambda D(A^*,\rho) +\delta D(\hat H,\rho)
\end{equation}
The Hamiltonian $H=-\epsilon Z/2$ is taken to be proportional to the third Paulli matrix $Z$. (This amounts to choosing a basis in the Hilbert space.)  $\epsilon>0$ is the energy gap of the two level system. $\mu\ge 0$ is the radiative decay rate, $\lambda\ge 0$ the excitation rate, and $\delta\ge 0$ the dephasing rate.
$D$ is a shorthand for
\begin{equation}
    D(B,\rho)= B\rho B^*- \frac 1 2 \{B^*B,\rho\}
\end{equation}
$A$ is the jump operator:
\begin{equation}\label{e:jump}
    A=\frac{X+iY}2=\kb{0}{1} ,\quad H\ket{0}=-\epsilon\ket{0}
\end{equation} 
$D(A,\rho)$ generates the radiative decay and $D(A^*,\rho)$ the excitation (by radiation or otherwise).  $D(\hat H,\rho)$ is a dephasing term 
\begin{equation}
    \hat H=\frac{H}{|H|}=-Z
\end{equation}
This two level systems is special in that $A$ and $H$ are simply related: 
 \begin{equation}\label{e:special}
    AH+HA=0
\end{equation}
In  section \ref{s:am} below we consider the situation where $A$ and $H$ are not simply related. 
The main result, Eq.~\ref{e:currents} below, is  independent of the relation between $A$ and $H$. 
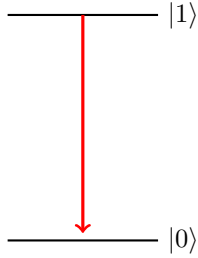
\begin{figure}[ht]
\begin{center}
    \begin{tikzpicture}
   \draw[thick] (-1,3) -- (1,3) node [right] {${\ket 1}$ };

   \draw[thick] (-1,0) -- (1,0) node [right] {$ {\ket 0} $};
  \draw[ very thick, red,->] (0,3) -- (0,.1);
    \end{tikzpicture}
 \caption{A radiating 2-level system.  
 }
\label{f:2level}
\end{center}
\end{figure}

The population rate of the ground state is given by differentiating Born rule
\begin{align}\label{e:rate}
    \frac{d}{dt}\bigg( \prob\Big(\kb{0}{0}\Big| \rho(t)\Big)\bigg)&=
    Tr \Big(\dot \rho \kb{0}{0}\Big)\nonumber \\ &= Tr\Big( L(\rho)\ \kb{0}{0}\Big)\nonumber \\ &= Tr\Big( \rho\  L^*(\kb{0}{0})\Big) ,
\end{align}
where 
\begin{equation}\label{e:Ls}
    L^*(M)=  {i}  [H,M] +\mu D^*(A,M) + \lambda D^*(A^*,M) +\delta D^*(Z,M),
\end{equation}
and
\begin{equation}\label{e:ds}
    D^*(B,M)= B^*M B- \frac 1 2 \{B^*B,M\} .
\end{equation}
$L^*(M)$ is the adjoint Lindbladian. It generates the Heisenberg evolution of observables:
\begin{equation}
    \dot M= L^*(M) .
\end{equation}
The population rate of the ground state is then $L^*(\kb{0}{0})$. The rhs of Eq.~\ref{e:rate} is therefore the Born rule for the population rate of the ground state.

Since the ground state $\kb{0}{0}$ commutes with $H$ and  $D^*(Z,\kb{0}{0})=0$,  the population rate is
\begin{equation}\label{e:currents}
      L^*(\kb{0}{0})=\mu D^*(A,\kb{0}{0}) + \lambda D^*(A^*,\kb{0}{0}) .
\end{equation}
The two terms on the right identify the radiation and excitation currents with the system observables $D^*(A,\kb{0}{0})$ and $D^*(A^*,\kb{0}{0})$. In particular,  the system observable corresponding to the radiation current is:
\begin{equation}\label{e:main}
\mu D^*(A,\kb{0}{0})=-\mu D^*(A,\kb{1}{1}) .
\end{equation}
Either side of Eq.~\ref{e:main} gives the system observable for the radiation current. The identity follows from
\begin{equation}
    D^*(\mathds{1})=0 .
\end{equation} 
 Similarly for the excitation currents. 

For the jump operator $A$ given in Eq.~\ref{e:jump}, one finds
\begin{equation}\label{e:spe}
\mu D^*(A,\kb{0}{0})=\mu\kb{1}{1}) .
\end{equation}
This gives Eq.~\ref{e:algorithm}.  Eq.~\ref{e:spe} is special as it depends on the simple relation between $A$ and $H$.   In the next section we look at the generic case.

\subsection{Broken rotational invariance}\label{s:am}

Dipole optical transitions occur between states that differ by (at most) one unit of angular momentum. The jump operators  are then of the form
\begin{equation}
A=\kb{{m'}}{m}, \quad m-m'\in 0, \pm 1 .  
\end{equation}
In the previous section we considered the case where the eigenstates of $H$ have well defined angular momenta. In the case that rotation invariance is broken, eigenstates of the energy are not eigenstates of angular momentum. In this case Eq.~\ref{e:main} does not reduce to Eq.~\ref{e:spe}. 

For the sake of concreteness consider a 3 level system spanned by states with definite angular momenta
\begin{equation}\label{e:amb}
    \ket{0}, \quad \ket{1},\quad \ket{2} .
\end{equation}
Suppose $\kb{0}{0}$ is the ground state with energy $0$ and  $\ket{1}$ and $\ket 2$ span the the two dimensional space of excited states which are almost degenerate near energy $1$.

\begin{figure}[ht]
\begin{center}
    \begin{tikzpicture}
  \draw[thick] (-1,3) -- (1,3) node [right] {$ 1 $ };
 \draw[thick] (-1,3.5) -- (1,3.5) node [right] {$ 2 $} ;
   \draw[thick] (-1,0) -- (1,0) node [right] {$ {0}$};
  \draw[ very thick, red,->] (0,3) -- (0,.1);
   \draw[ very thick, red,->] (0,3.5) -- (0,3.1);
    \end{tikzpicture}
 \caption{A radiating 3-level system where the  excited states are hybridization of different angular momentum states. $j $ denote eigenstates of energy.
 }
\label{f:3level}
\end{center}
\end{figure}
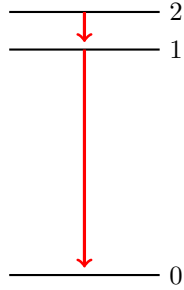

The two jump operators are 
\begin{equation}
    A_{1\to 0}= \kb{1}{0}, \quad A_{2\to 1}=\kb{1}{2} .
\end{equation}
For the Hamiltonian choose
\begin{equation}
    H= \Big(\kb{1}{1}+\kb{2}{2}\Big)+ \epsilon\Big(\kb{1}{2}+\kb{2}{1}\Big) ,
\end{equation}
with $1\gg\epsilon>0$. The Hamiltonian fails to commute with the angular momentum by $\epsilon$. However, because the excited states are almost degenerate, the eignestates of energy are far from eigenstates of angular momentum. In particular, the projection on the first excited state $P_1$,  is given by
\begin{equation}
    P_1= \frac 1 2 \Big( \kb{1}{1}+\kb{2}{2}-\kb{1}{2}-\kb{1}{2}\Big), \quad HP_1=(1-\epsilon)P_1 \  .
\end{equation}
This is an equal superposition of different angular momentum states.

The radiation currents can be represented either in the angular momentum basis, Eq.~\ref{e:amb}, or in the energy basis which we denote by a  round bracket
\begin{equation}
    |0)=\ket{0}, \quad |1)=\frac{\ket 1-\ket 2}{\sqrt 2}, \quad |2)=\frac{\ket 1+\ket 2}{\sqrt 2} .
\end{equation}

The radiation current $1\to 0$ is given, in the angular momentum basis, by
\begin{equation}\label{e:s}
    \mu_{1\to 0} D^*(A_{1\to 0},\kb{0}{0})=\mu_{1\to 0} \Big(\kb{1}{1}\Big) ,
\end{equation}
and in the energy basis by
\begin{equation}
    \frac{\mu_{1\to 0}}{2} 
     \Big(\kbp{1}{1}+\kbp{2}{2}+  \kbp{1}{2}+\kbp{2}{1}\Big)_{\text{energy basis}} .
\end{equation}
The radiation current associated with the $2\to 1$ decay is given by
\begin{equation}
    \mu_{2\to 1} D^*(A_{2\to 1},P_1)=   \frac{ \mu_{2\to 1}}{4} \Big(  \kb{1}{2}+\kb{2}{1}\Big) ,
\end{equation}
in the angular momentum basis, and in the energy basis by 
\begin{equation}\label{e:f}
       \frac{ \mu_{2\to 1}}{4} \Big(  \kbp{1}{1}-\kbp{2}{2}\Big)_{\text{energy basis}} 
\end{equation}
It is probably a  challenge to recover Eqs.~\ref{e:s}-\ref{e:f} without using Eq.~\ref{e:main} \footnote{We found it convenient to use  Mathematica.}.


\section{Heralding is consistent with no cloning}

The ability to prepare a quantum system remotely is the discovery of EPR \cite{einstein1935}. It is the standard way of preparing a radiating system, also known as heralding \cite{heralding}.  Heralding is an important resource of quantum technologies. 

In the simplest heralding scheme, Alice and Bob start with a shared maximally entangle (Bell) state. By testing her qubit, Alice prepares her qubit in the state $\ket\psi_A$, and prepares Bob's qubit in a "mirror" state $\ket{\bar\psi}$ (see Eq.~\ref{e:mirror} below). Since this creates two "mirror" images of the a-priori arbitrary state $\ket\psi_A$, one wonders if heralding violates  no-cloning \cite{clone}. 

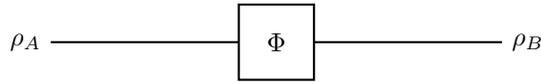
\begin{figure}[ht]
\begin{center}
    \begin{tikzpicture}
    \begin{scope}
   \draw[thick] (-1/2,-1/2) rectangle (1/2,1/2);
    \draw[thick] (-3,0) -- (-1/2,0)  node [left] at (-3,0) {$\rho_A$} node [right] at (3,0) {$\rho_B$};
   \draw[thick] (1/2,0) -- (3,0);
    \node at (0,0) {$\Phi$};
    \end{scope}
        \end{tikzpicture}
 \caption{A quantum channel}
\label{f:choi}
\end{center}
\end{figure}

The basic primitive for transmission of quantum information is a quantum channel \cite{holevo, wilde}:
\begin{equation}
    \rho_B=\Phi(\rho_A) .
\end{equation}
$\Phi$ is, by definition, a completely positive trace preserving linear map (CPTP)  \cite{holevo, nielsen}. 
Herlading is not a completely positive map and  hence can not be interpreted as a quantum channel. We shall use this fact to reconcile heralding with no-cloning. 

Consider a Bell state shared by Alice and Bob
\begin{equation}
    \sqrt 2 \ket\beta_{AB}=\ket{00}_{AB}+\ket{11}_{AB} .
\end{equation}
Alice makes a successful test of
\begin{equation}
    \ket\psi_A=\psi_0\ket 0_A+\psi_1\ket 1_A .
\end{equation}
This heralds Bob the ``mirror" state
\begin{equation}
    \sqrt 2\braket{\psi}{\beta}_B=\bar\psi_0\ket 0_B+\bar\psi_1\ket 1_B=\ket{\bar\psi}_B ,
\end{equation}
where bar denotes complex conjugation. The mapping of the Alice's vector to Bob's vector
\begin{equation}\label{e:mirror}
     \ket\psi_A\mapsto \ket{\bar\psi}_B  
\end{equation}
translates to mapping of Alice's state to Bob's state:
\begin{equation}
     \Big(\ket\psi\!\bra\psi\Big)_A=\begin{pmatrix}|\psi_0|^2&\psi_0\bar\psi_1\\
     \psi_1\bar\psi_0& |\psi_1|^2
     \end{pmatrix}_A
     \mapsto \Big(\ket{\bar\psi}\!\bra{\bar\psi}\Big)_B=\begin{pmatrix}|\psi_0|^2&\bar\psi_0\psi_1\\
     \bar\psi_1\psi_0& |\psi_1|^2
     \end{pmatrix}_B \ .
\end{equation}
The map $\Phi$ is transposition
\begin{equation}
  \rho_B=  \Phi(\rho_A)= \rho^T_A\ .
\end{equation}
Transposition acts on the Bloch ball as reflection $Y\mapsto -Y$. It is  the canonical example of positive map that is not completely positive \cite{nielsen, wilde}. 

There is no unitary, independent of  $
\ket\psi_A$, that Bob can apply to his state to recover Alice's state. This can be seen by observing that reflection  does not preserve the Pauli operators commutation relations. Therefore, it can't be implemented as a unitary\footnote{It can be implemented by an anti-unitary.}.   Bob can not make a precise copy of $\ket\psi_A$ without additional information.  
Of course, if Alice gives Bob classical information about her test $\ket\psi_A$, for example,  the protocol for making $\ket\psi_A$, Bob could (probabilistically) reproduce $\ket\psi_A$ without using the available entanglement resource.

The inability of Bob to recover Alice's state is a health certificate. For, if Bob could use heralding to create a copy of $\ket\psi_A$, this would be a trap-door to cloning  \cite{clone}. 



\section*{Acknowledgement} JEA thanks Gilad Gour and Yoav Sagi for helpful comments.

\appendix
\section{Lindblad evolution of a two level system}
 
The Lindbladian in Eq.~\ref{e:lind} admits an elementary solution of the equations of motion.  Write $\rho\ge 0$ as
\begin{equation}
    \rho=\frac{\mathds{1}+x X+yY+zZ}{2}, \quad x^2+y^2+z^2\le 1
\end{equation}
with $X,Y,Z$ the three Pauli matrices. The Lindblad equation of motion  takes the from
\begin{align}
    \dot x&=-\beta\, x-\epsilon y, \nonumber \\
    \dot y&= \epsilon x-\beta y, \\
     \dot z&=-\lambda+\mu -(\lambda+\mu)\, z \nonumber
\end{align}
with 
\begin{equation}
     \beta=\frac{2\delta+\lambda+\mu}{2}\ge 0
\end{equation}
The motion along the z axis describes exponential relaxation to the stationary value $z_\infty$
\begin{equation}
    z(t)= z_0 e^{-(\lambda+\mu)t}+z_\infty, \quad  z_\infty=\frac{\lambda-\mu}{\lambda+\mu}
\end{equation}
The equation of  motion in the x-y plane can be combined to
\begin{equation}
    \dot \zeta=(i\epsilon -\beta)\, \zeta, \quad \zeta=x+iy
\end{equation}
whose solution describes an exponential spiral towards the axis
\begin{equation}
    \zeta(t)=\zeta_0 e^{(i\epsilon-\beta))t}\ .
\end{equation}

\printbibliography 

\end{document}